\documentstyle[11pt,paspconf,epsf]{article}

\begin{document}

\title{Model atmosphere calibration for synthetic photometry applications :
       the $uvby$ Str\"omgren photometric system}

\author{Th. Lejeune}
\affil{Astronomisches Institut der Universit\"at Basel, Venusstr. 7, 
    CH-4102 Binningen, Switzerland}
    
\author{E. Lastennet}
\affil{Astronomy Unit, Queen Mary and Westfield College,
    Mile End Road, London E1 4NS, UK}
    
\author{P. Westera, and R. Buser}
\affil{Astronomisches Institut der Universit\"at Basel, Venusstr. 7, 
    CH-4102 Binningen, Switzerland}

\begin{abstract}
Str{\"o}mgren synthetic photometry from an empirically calibrated grid
of stellar  atmosphere  models has been used  to  derive the effective
temperature of each   component  of double  lined  spectroscopic (SB2)
eclipsing  binaries.  For this purpose,  we have selected a sub-sample
of 20 SB2s for which (b$-$y), m$_1$, and  c$_1$ individual indices are
available.  This new determination  of effective temperature  has been
performed in a homogeneous  way for all these  stars. As the effective
temperature determination  is  related to the assumed  metallicity, we
explore simultaneous solutions in the ($T_{\rm eff}$,[Fe/H])-plane and
present   our results as  confidence   regions  computed to match  the
observed values of surface gravity,  (b$-$y), m$_1$, and c$_1$, taking
into  account   interstellar   reddening.  These   confidence regions,
presented   in  detail in Lastennet  et   al. 1999, show that previous
estimates of $T_{\rm eff}$ are  often too optimistic, and that  [Fe/H]
should not be   neglected  in such  determinations. We   present  some
comparisons with Ribas et al. (1998) using  Hipparcos parallaxes for 8
binaries of  our working sample, showing good  agreement with the most
reliable  parallaxes. This  point gives  a significant  weight to  the
validity of the BaSeL models for synthetic photometry applications.
\end{abstract}



\section{Introduction}

Since predictions  based on stellar-atmosphere  models  are useful for
{\it ``Spectro-photometric    dating of  stars  and  galaxies"},  main
subject  of this conference, we  present some of  the results obtained
from an  empirically calibrated grid of  stellar atmosphere models for
simultaneously    deriving    homogeneous  effective  temperatures and
metallicities of 40 stars from observed data.

\section{BaSeL models} 

We use  the Basel  Stellar  Library (BaSeL)  photometric calibrations,
extensively  tested and  regularly  updated   for   a larger set    of
parameters  (see   Lejeune et  al.   1997,  1998   and  Lastennet   et
al.  1999).  The  BaSeL  models cover  a   large range  of fundamental
parameters: 2000 K $\leq$   $T_{\rm  eff}$ $\leq$ 50,000  K,   $-$1.02
$\leq$ $\log   g$ $\leq$ 5.5,  and  $-$5.0 $\leq$ [Fe/H]  $\leq$ +1.0.
This library combines  theoretical stellar energy  distributions which
are  based on several original  grids  of blanketed model atmospheres,
and which have been  corrected in such  a way as to provide  synthetic
colours  consistent  with  extant    empirical calibrations   at   all
wavelengths from the near-UV through the far-IR.  For more details and
references on the BaSeL library,  see contributions of Lejeune et al.,
Lastennet et al. and Westera et al. in this volume.

\section{Comparison with Hipparcos parallax}

Very  recently,  Ribas et  al.    (1998) have  computed the  effective
temperatures of  19 eclipsing  binaries    included in the   Hipparcos
catalogue from their   radii, Hipparcos trigonometric  parallaxes, and
apparent visual   magnitudes  corrected  for   absorption.   They used
Flower's (1996) calibration to  derive bolometric corrections.  Only 8
systems are in common with our working sample. The comparison with our
results is made in  Table 1.  The $T_{\rm  eff}$ being  highly related
with metallicity, a direct comparison  is not possible because, unlike
the  Hipparcos-derived data, our  results  are not  given in terms  of
temperatures with    error  bars, but  as   ranges   of $T_{\rm  eff}$
compatible with  a given [Fe/H].  Thus, the  ranges reported in Tab. 1
are given assuming  three different hypotheses:  [Fe/H]$=-$0.2, [Fe/H]
$=$ 0,  and   [Fe/H] $=$   0.2.   The   overall  agreement is    quite
satisfactory,  as illustrated  in Fig. 1.   The  disagreement for  the
temperatures of CW Cephei  can be explained  by the large error of the
Hipparcos parallax ($\sigma$$_{\rm  \pi}$/$\pi$  $\simeq$70\%).    For
such  large errors, the Lutz-Kelker correction   (Lutz \& Kelker 1973)
cannot be neglected: the  average distance is certainly underestimated
and,  as a  consequence, the  $T_{\rm  eff}$ is also underestimated in
Ribas  et al.'s (1998)  calculation.   Thus,  the  agreement with  the
results obtained from  the BaSeL  models is  certainly better than  it
would appear in Fig.  1 and Tab.  1.  Similar corrections, of slightly
lesser extent, are probably also indicated for the $T_{\rm eff}$ of RZ
Cha and GG Lup, which have $\sigma$$_{\rm \pi}$/$\pi >$ 10{\%} (11.6\%
and   11.4\%, respectively).  Finally, it   is  worth noting that  the
system with the smallest relative error in Tab.  1, $\beta$ Aur, shows
excellent agreement   between $T_{\rm  eff}$  (Hipparcos) and  $T_{\rm
eff}$ (BaSeL), which underlines the validity of the BaSeL models.

\begin{figure}
\plotfiddle{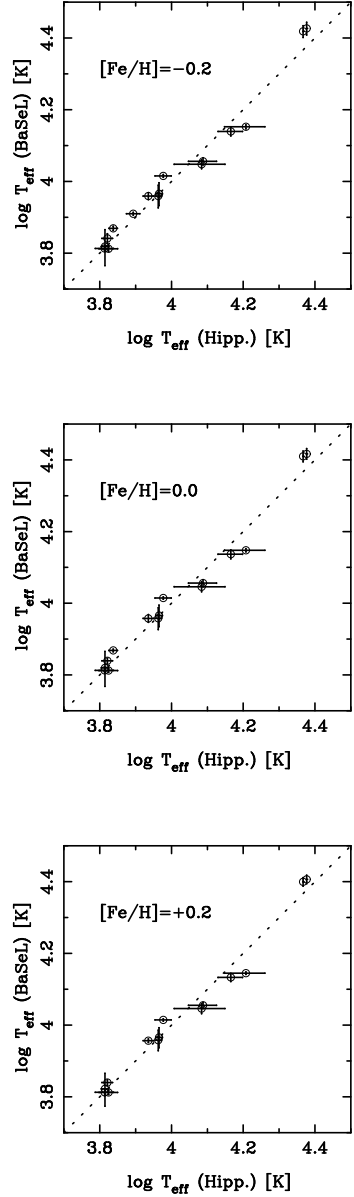}{15. cm}{-90}{85}{85}{-360}{500}
 \caption{Hipparcos- versus   BaSeL-derived effective temperatures for
 $\beta$ Aur,  YZ Cas, CW Cep,  RZ  Cha, KW Hya,  GG Lup,  TZ Men, and
 $\zeta$ Phe.  The errors  are not shown on the  Hipparcos axis for CW
 Cephei  (the hottest binary     in  these figures).  See   text   for
 explanation.}
\end{figure}


\begin{table}
	\caption[]{Effective temperatures  from Hipparcos (after Ribas
	et al. 1998) and from BaSeL models matching (b$-$y)$_0$, m$_0$,
	c$_0$,  and $\log g$ for the  three   following metallicities:
	[Fe/H]$ =-$0.2, 0 and 0.2.} 
\label{tab:Hipp}
\vspace{0.2cm}
\scriptsize
\begin{tabular}{lrrcrcrc}
\tableline\noalign{\smallskip}
Name         &       &      \multicolumn{2}{c}{[Fe/H]$=-$0.2}        &
\multicolumn{2}{c}{[Fe/H] $=$ 0.} & \multicolumn{2}{c}{[Fe/H] $=$ 0.2} \\
\noalign{\smallskip}
\tableline
\noalign{\smallskip}
     &  $T_{\rm eff}$(Hipp.)  [K]   &   $T_{\rm eff}$(BaSeL) [K]   &  $\sigma$  &
     $T_{\rm eff}$(BaSeL) [K] &  $\sigma$ & $T_{\rm eff}$(BaSeL) [K] & $\sigma$
     \\
\noalign{\smallskip}
\tableline
\noalign{\smallskip}
$\beta$ Aur & 9230$\pm$150   & [8780,9620]   & 1 & [8780,9560]   & 1 & [8900,9500]   & 1 \\
            & 9186$\pm$145   & [8540,9500]   & 1 & [8600,9440]   & 1 & [8660,9320]   & 1 \\
YZ Cas      & 8624$\pm$290   & [9000,9120]   & 2 & [8920,9240]   & 3 & no solution   &   \\
            & 6528$\pm$155   & [6100,7140]   & 1 & [6180,7060]   & 1 & [6260,7060]   & 1 \\
CW Cep      & 23804          & [26000,27200] & 1 & [25600,26600] & 1 & [24600,26600] & 2 \\
            & 23272          & [25600,26800] & 1 & [25200 26200] & 1 & [24800,25400] & 1 \\
RZ Cha      & 6681$\pm$400   & [6440,6560]   & 1 & [6380,6600]   & 2 & [6340,6640]   & 3 \\ 
            & 6513$\pm$385   & [6420,6580]   & 1 & [6460,6540]   & 1 & [6420,6580]   & 2 \\
KW Hya      & 7826$\pm$340   & [8080,8100]   & 3 & no solution   &   & no solution   &   \\
            & 6626$\pm$230   & [6780,7120]   & 3 & [6860,6980]   & 1 & [6860,7000]   & 3 \\
GG Lup      & 16128$\pm$2080 & [14080,14260] & 1 & [14020,14140] & 1 & [13780,14140] & 2 \\
            & 12129$\pm$1960 & [10920,11320] & 1 & [10920,11320] & 1 & [10920,11320] & 1 \\
TZ Men      & 9489$\pm$490   & [10300,10420] & 1 & [10300,10380] & 1 & [10260,10460] & 2 \\
            & 6880$\pm$190   & [7340,7460]   & 3 & no solution   &   & no solution   &   \\
$\zeta$ Phe & 14631$\pm$1150 & [13540,14020] & 1 & [13460,13860] & 1 & [13380,13860] & 1 \\
            & 12249$\pm$1100 & [11240,11560] & 1 & [11280,11480] & 1 & [11040,11680] & 2 \\
\noalign{\smallskip}
\tableline
\tableline
\end{tabular}
\end{table}

\normalsize

\section{Brief summary of the results}
 
\begin{itemize}
\item 
The large range of [Fe/H] associated with acceptable confidence levels
makes  it evident  that the  classical method to  derive $T_{\rm eff}$
from metallicity-independent  calibrations should be  considered  with
caution.
\item 
By exploring  the best $\chi^2$-fits to the  photometric data, we have
re-derived new reddening values for some stars.
\item 
Comparisons  for   16   stars   with Hipparcos-based   $T_{\rm   eff}$
determinations show good agreement  with the temperatures derived from
the BaSeL models. The agreement is  even excellent for the star having
the most reliable Hipparcos data in the sample studied.
\end{itemize}

See Lastennet  et   al. 1999   for   details about  the method,    the
determination of reddening, the influence  of gravity, etc... \\ These
comparisons also  demonstrate   that, while originally  calibrated  in
order  to reproduce the broad-band    (UBVRIJHKL) colours, the   BaSeL
models also provide  reliable results for  medium-band photometry such
as the Str{\"o}mgren photometry. This point gives a significant weight
to   the validity  of the   BaSeL  library  for synthetic   photometry
applications in general.

\acknowledgments
T.   L. gratefully  acknowledges  financial  support  from the   Swiss
National  Science Foundation  (grant 20-53660.98 to  Prof. Buser), and
from   the  ``Soci\'et\'e  Suisse  d'Astronomie   et d'Astrophysique''
(SSAA). We  are  grateful  to the  organisers  for arranging  such  an
enjoyable meeting.

\end{document}